\documentclass[aps,prl,preprint,superscriptaddress,showpacs]{revtex4-1}

\usepackage{graphicx}% Include figure files
\usepackage{dcolumn}% Align table columns on decimal point
\usepackage{bm}% bold math
\usepackage{amsmath}
\usepackage{ulem}
\usepackage[caption=false]{subfig}
\usepackage{color}

\newcommand{\overbar}[1]{\mkern 1.5mu\overline{\mkern-1.5mu#1\mkern-1.5mu}\mkern 1.5mu}

\begin{document}

%\preprint{}

\title{Spin waves in the anisotropic fcc kagome antiferromagnet}


%\author{}
%\thanks{}
%\affiliation{}
\author{M. D. LeBlanc}
%\thanks{}
\affiliation{Department of Physics and Physical Oceanography, Memorial University of Newfoundland, St. John's, Newfoundland, A1B 3X7,  Canada}

\author{B. W. Southern}
%\thanks{}
\affiliation{Department of Physics and Astronomy, University of Manitoba, Winnipeg, MB, R3T 2N2, Canada}
\author{M. L. Plumer}
%\thanks{}
\affiliation{Department of Physics and Physical Oceanography, Memorial University of Newfoundland, St. John's, Newfoundland, A1B 3X7,  Canada}
\author{J. P. Whitehead}
%\thanks{}
\affiliation{Department of Physics and Physical Oceanography, Memorial University of Newfoundland, St. John's, Newfoundland, A1B 3X7,  Canada}


\date{\today}

\begin{abstract}
Spin wave calculations demonstrate that the macroscopic continuous degeneracy associated with the two-dimensional kagome Heisenberg spin lattice persists in the case of the stacked fcc structure giving rise to zero energy modes in three dimensions.  The addition of an effective local cubic anisotropy is shown to remove this continuous degeneracy and introduce a gap in the spectrum as well as modify the inelastic scattering function $S({\bf q},\omega)$.  This scenario supports earlier Monte Carlo simulations which indicate that the phase transition to long range $q=0$ magnetic order is driven to be discontinuous by critical fluctuations associated with the large degeneracy in the absence of anisotropy, but becomes continuous with the addition of anisotropy. The results are relevant to Ir-Mn alloys which are widely used in the magnetic storage industry in thin film form as the antiferromagnetic pinning layer in spin-valve structures.
\end{abstract}

%\keywords{}

\maketitle

%\section{INTRODUCTION}

The Heisenberg model on an isolated 2D kagome layer with nearest neighbour antiferromagnetic exchange interactions is an example of a fully frustrated magnetic system with a macroscopically degenerate ground state of 120$^0$ spin structures on corner-sharing triangles, known as $q=0$ magnetic order \cite{chalker1992}. This extensive degeneracy is accompanied by the presence of zero energy spin wave excitations which can take the system from one ground state to another \cite{harris1992,schnabel2012}. Evidence for such a mode through inelastic neutron scattering experiments has been reported in a system with weakly coupled kagome layers \cite{matan06}. Zero-energy modes were also predicted to occur in the stacked triangular lattice antiferromagnet where the inter-layer exchange coupling $J'$ differs from the intra-layer interaction $J$ (a model of the magnetism in solid oxygen) \cite{rastelli86,jansen86}. In this case of rhombohedral symmetry, the degeneracy is associated with ground state helimagnetism and occurs if $|J'| < 3|J|$. For this system, degenerate modes occur along lines in reciprocal space that are dependent on the value of $J'$. Similar macroscopic degeneracies are found in spin ice materials \cite{bramwell2001,zhitomirsky2012,wong2013,ross14} and can often be lifted by thermal or quantum fluctuations through the mechanism of order by disorder in which states are selected from the ground state manifold by entropic forces \cite{villain80,henley89}. Such degeneracies can also be removed with the addition of further neighbour interactions or magnetic anisotropies \cite{rastelli86,matan06,shahbazi08}.

\begin{figure}[b]
 \includegraphics[width=3in]{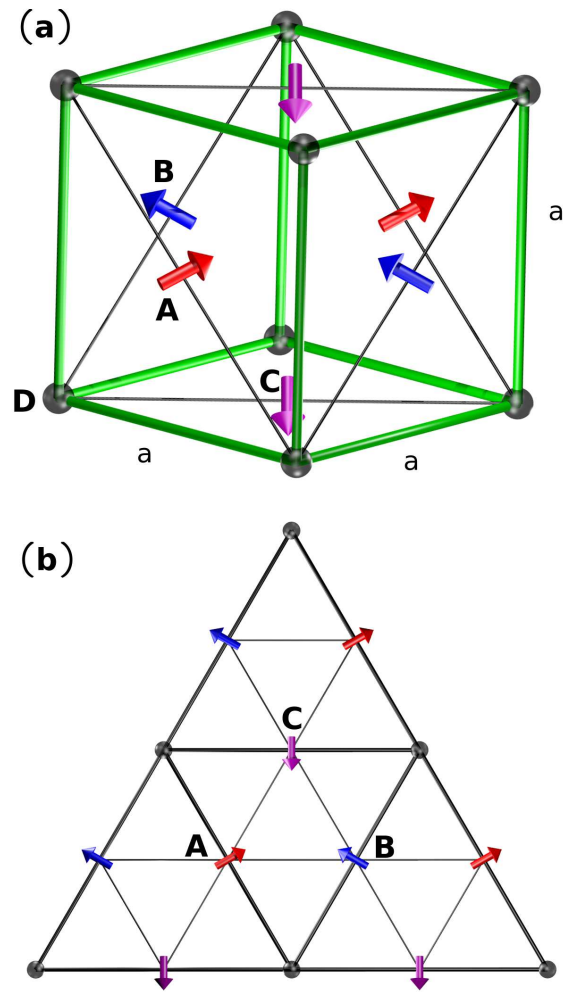} 
 \caption{(a) The fcc lattice is divided into four cubic sublattices, each with lattice constant a and labelled as ${\rm A,B,C}$, and ${\rm D}$. The ${\rm D}$ sites (spheres) are non-magnetic whereas the remaining three sublattices (arrows) are magnetic. (b) The ${\rm A,B}$, and ${\rm C}$ sites form a set of kagome lattices stacked along the (111) direction. Figure adapted from Ref.~\cite{tomeno1999}.}\label{fig1}
 \end{figure}

Monte Carlo simulations have recently been reported on 3D kagome Heisenberg and XY spin systems composed of stacked layers with fcc symmetry (see Fig.~\ref{fig1}) \cite{hemmati2012}. The results suggest that continuous macroscopic degeneracy associated with the 2D system persists in this 3D spin lattice. The degeneracy in the 3D system does not yield a finite ground state entropy, but the degeneracy scales with the linear size of the system. In the absence of anisotropy, the 3D system exhibits a finite temperature phase transition that is weakly first order, speculated to be driven so by critical fluctuations associated with the large degree of degeneracy. With the addition of a local cubic anisotropy \cite{leblanc2013,szunyogh2009}, the nature of the transition becomes continuous, a phenomenon also reported in spin-ice materials \cite{zhitomirsky2012}.

The magnetic properties of Ir-Mn alloys have attracted a great deal of interest due to their widespread use as the antiferromagnetic pinning layer in spin valve devices for magnetic recording \cite{berk1999,ogrady2010,tsunoda10}.  Ordered IrMn$_3$, and sister compounds RhMn$_3$ and PtMn$_3$, adopt the CuAu$_3$ crystal structure\cite{kren66} with magnetic Mn ions on cube faces and non-magnetic (Ir) occupying cube corners.  The high N\'{e}el temperature of IrMn$_3$ ($T_N$ = 960 K) makes it particularly useful for applications that require robustness to thermal fluctuations at device operating temperatures.  Despite extensive study of their magnetic structures spanning many decades, including a seminal neutron diffraction work identifying the 120$^0$ spin structure \cite{tomeno1999}, the connection to the kagome structure was not realized until recently \cite{hemmati2012}. This has inspired a new study of IrMn$_3$ as possible candidates for observing the anomalous Hall effect in zero applied magnetic field \cite{chen2014}.  A recent neutron diffraction study of single-crystal IrMn$_3$ shows that the $q=0$ spin structure remains for thin films and emphasizes the importance of this frustrated magnetic ordering on the exchange bias phenomena in this compund \cite{kohn2013}. This relationship, and the importance of anisotropy, has also been revealed in recent electronic structure calculations and micromagnetic simulations on IrMn$_3$/Co bilayer thin films \cite{yanes2013}.

In the present work we examine the spin wave excitations as well as the scattering intensity function $S({\bf q},\omega)$ of a model Hamiltonian with unequal intra-plane and inter-plane antiferromagnetic exchange interactions including local cubic anisotropy. It is shown that both the addition of inter-plane exchange and anisotropy either reduce or remove the macroscopic degeneracy associated with the 2D kagome system and that this behaviour is related to the reduction or removal of the number of zero energy modes in the excitation spectrum. The results provide guidance for the potential observation of these effects through inelastic neutron diffraction measurements of IrMn$_3$ and its sister compounds.

%\section{The Model}

The magnetic ions form an fcc lattice with one of the four cubic sublattices replaced by non-magnetic sites as shown in Fig.~\ref{fig1}. The three remaining sublatttices form stacked kagome layers parallel to the [111] directions and are labelled as A, B and C. Each magnetic site has four nearest neighbours (NN) in the (111) planes and two NN in each of the planes above and below. We consider NN exchange interactions as well as a local cubic anisotropy described by the Hamiltonian
\begin{equation}
 {\cal H} = J\sum_{i<j}^{\rm intraplane} \mathbf{S}_i \cdot \mathbf{S}_j + J'\sum_{i<j}^{\rm interplane} \mathbf{S}_i \cdot \mathbf{S}_j- K\sum_{\gamma}\sum_{l\subset \gamma} (\mathbf{S}_l \cdot \mathbf{n}_{\gamma})^2  
\end{equation}
where  $J>0$ is the antiferromagnetic coupling to the four in-plane NNs and $J' \ge 0$ couples the four out-of-plane NN spins, and the anisotropy  $K \ge 0$ has a different easy direction for each of the three sublattices. Here,  $\gamma$ represents sublattice A, B and C and $l$ is summed over the $\frac{N}{3}$ spins of sublattice $\gamma$, $\mathbf{S}_i$ are unit classical Heisenberg spin vectors at each site and  $\mathbf{n}_\gamma$ are unit vectors in the cube axes directions, $\mathbf{n}_{\rm A}=\hat{\bf{x}}$, $\mathbf{n}_{\rm B}=\hat{\bf{y}},$ and $\mathbf{n}_{\rm C}=\hat{\bf{z}}$.   Electronic structure calculations \cite{szunyogh2009} have been used to estimate  $K/J$ $\approx 0.1$ in the case of IrMn$_3$. 

For zero anisotropy, the ground state is a planar configuration with the sum of the spins on each elementary triangle $\mathbf{S}_A+\mathbf{S}_B+\mathbf{S}_C=0$. For decoupled kagome planes ($J'=0$) there is an extensive ground state degeneracy  which includes both periodic and non-periodic ground states. Previous spin wave calculations at zero temperature for the $q=0$ and $\sqrt{3} \times \sqrt{3}$ periodic ground states find zero energy dispersionless modes which are related to the macroscopic degeneracy \cite{harris1992}. The  decoupled kagome planes do not order at a finite temperature but are believed to select a planar arrangement of the spins as $T \rightarrow 0$ and spin dynamic simulations indicate strong $\sqrt{3} \times \sqrt{3}$ spin correlations at low T \cite{reimers1993,schnabel2012}.  For coupled planes ($J' > 0$), there is a finite temperature phase transition to the $q=0$ state which is weakly first order \cite{hemmati2012}. The ground state degeneracy is no longer extensive but there are continuous rotations of two of the three sublattices which do not change the energy and correspond to local modes. When the cubic anisotropy is added, particular spin planes are selected and all of the macroscopic degeneracy is removed with the phase transition changing from first order to second order. There are eight possible ground states corresponding to the (111) planes with the spins being lifted out of the plane resulting in a net magnetization along one of the [111] axes \cite{leblanc2013}.

%\section{Spin Waves}
In order to study the linearized spin wave excitations, we consider a single domain in which the net magnetization is along the positive [111] direction. The spins $\mathbf{S}_i$ on each sublattice are transformed to local spin coordinates $\mathbf{\tilde{S}}_i$ such that $\tilde{S}_i^z = 1$ in the ground state. We look for plane wave solutions involving the transverse spin components $\tilde{\bf{S}}_i = \tilde{\bf{S}} e^{i(\mathbf{k}\cdot\mathbf{r}_i -\omega t)}$ and the linearized equations for the six transverse spin amplitudes can be obtained through the standard torque equation \cite{torque} or other techniques \cite{morra1988}.

%\subsection{Zero Anisotropy}
The linearized equations yield real eigenvalues $\pm \omega_1, \pm \omega_2, \pm \omega_3$. In the general case, these values must be obtained numerically but analytic results can be determined in special cases. For $K=0$, the ground state is the planar spin configuration with the three sublattices oriented at 120$^0$ with respect to each other. In this case the problem can be reduced to finding the eigenvalues $\omega^2$ of a $ 3 \times 3$ symmetric matrix.
If the interplane coupling $J'$ is also zero,   
the characteristic cubic equation has a zero eigenvalue for all $\mathbf{k}$ and the remaining two eigenvalues are degenerate and given by the following expressions
\begin{equation}
\begin{split}
\omega_1 =&\ 0 \\
\omega_{2,3} =&\ \sqrt{2}J \{\sin[(k_x-k_y)a/2]^2 +\sin[(k_x-k_z)a/2]^2 \\
            & \qquad\ \ + \sin[(k_y-k_z)a/2]^2\}^{1/2}
\end{split}
\label{eq:simple}
\end{equation}
where the wavevector components $k_x, k_y, k_z$ are defined with respect to the cubic axes with lattice constant $a$.  These expressions agree with previous results \cite{harris1992} for the NN $q=0$ kagome spin lattice when $a=\sqrt{2}$ (corresponding to a NN distance of unity). The dispersionless mode is related to the local rotations of the spins from one ground state to another. Note that for $k_x=k_y=k_z$ all three modes are dispersionless. The latter case corresponds to the fact that the decoupled kagome planes can have arbitrary uniform rotations with respect to one another. For $\mathbf{k}$ along one of the cube axes, Eq.~\ref{eq:simple} reduces to $\omega_2=\omega_3= 2 J |\sin(ka/2)|$.

When the interplane interaction $J' > 0$, all three modes are dispersive and non-degenerate in general. There are two special cases where degeneracy occurs and where zero frequency modes are present. For $k_x=k_y=k_z$, analysis shows that
\begin{equation}
\begin{split}
 \omega_1^2=\omega_3^2=&\ [1-\cos(k_xa)][4J'^2+6JJ'+2J'^2 \cos(k_xa)] \\
 \omega_2^2=&\ [1-\cos(k_xa)][4J'^2+12JJ'+8J'^2\cos(k_xa)]
\end{split}
\end{equation}
All three modes are dispersive due to the coupling between kagome planes and two are degenerate. However, note that $\omega_2$ becomes a soft mode at the zone boundary $k_x=\pi/a$ when $J'=3J$. For values of the inter-plane coupling $J' > 3J$ the ground state is no longer the $q=0$ kagome state but rather corresponds to ferromagnetic kagome planes which are ordered antiferromagnetically with respect to each other. In this paper we restrict our considerations to $J' < 3J$.

In the second special case, $k_y=k_z=0$, which corresponds to spin waves propagating parallel to one of the cubic crystal axes, we have 
\begin{equation}
\begin{split}
\omega_1 =&\ 0 \\
\omega_2 = \omega_3 =&\ 2(J+J')|\sin(k a/2)|
\end{split}
\end{equation}
where $k$ represents $k_x, k_y$, or $k_z$. These expressions reduce to the correct 2D result above when $J'=0$. Hence, the coupling of the planes stiffens the excitations for wavevectors along the crystal axes but does not remove the zero frequency mode. 

The zero mode can be understood from Fig.~\ref{fig1}. The $x=na$ planes only have B sites whereas the $x=(n+1/2)a$ planes have both A and C sites where $n=0,1,2,...$. In the ground state the A and C sublattices are at 120$^0$ to each other and to the B sublattice. The entire plane of AC spins can be  rotated continuously about the direction of the B sublattice spins in the planes on either side with no change in energy. In addition, these rotations in each of the AC planes are independent and correspond to a set of localized excitations for $J+J' >0$. When $J'=0$, there are additional degeneracies which lead to a zero mode for all $\mathbf{k}$ \cite{harris1992}.

%\subsection{Effects of Anisotropy}
\begin{figure}[ht]
\centering
\subfloat{\label{subfig:J0K0}\includegraphics[width=60mm,height=50mm]{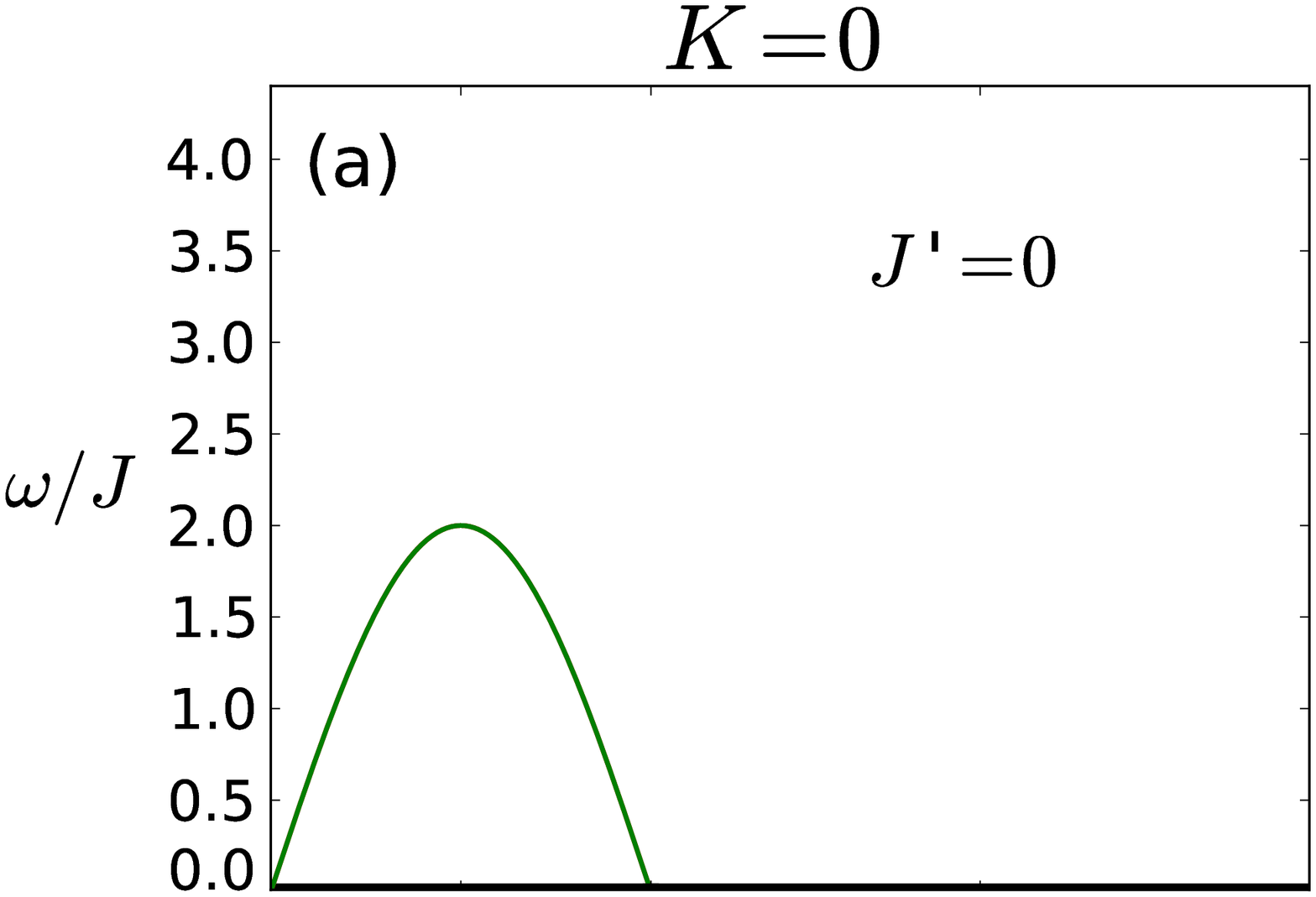}}
\subfloat{\label{subfig:J0K0.1}\includegraphics[width=50mm,height=50mm]{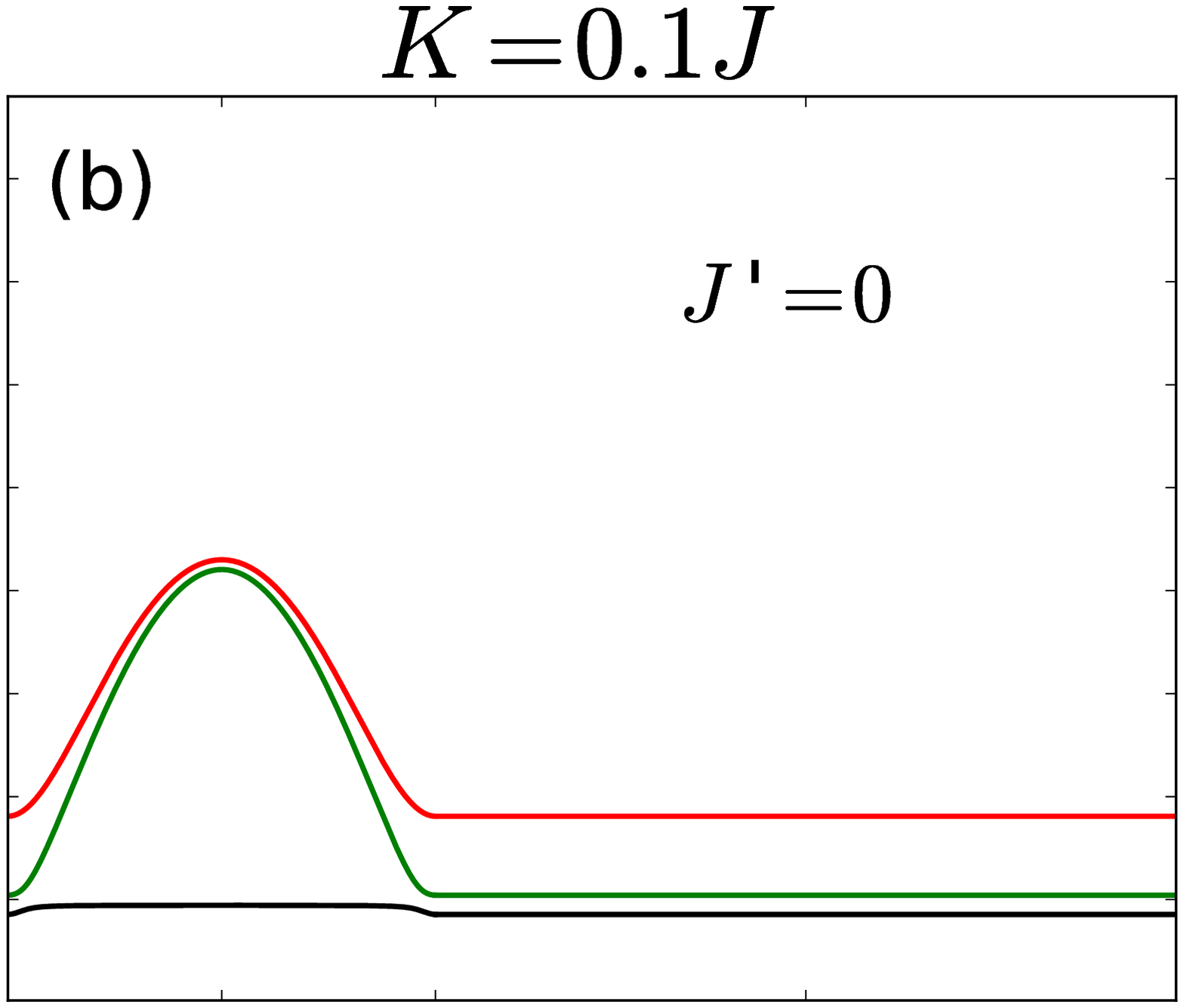}}
\\
\subfloat{\label{subfig:J0.1K0}\includegraphics[width=60mm,height=45mm]{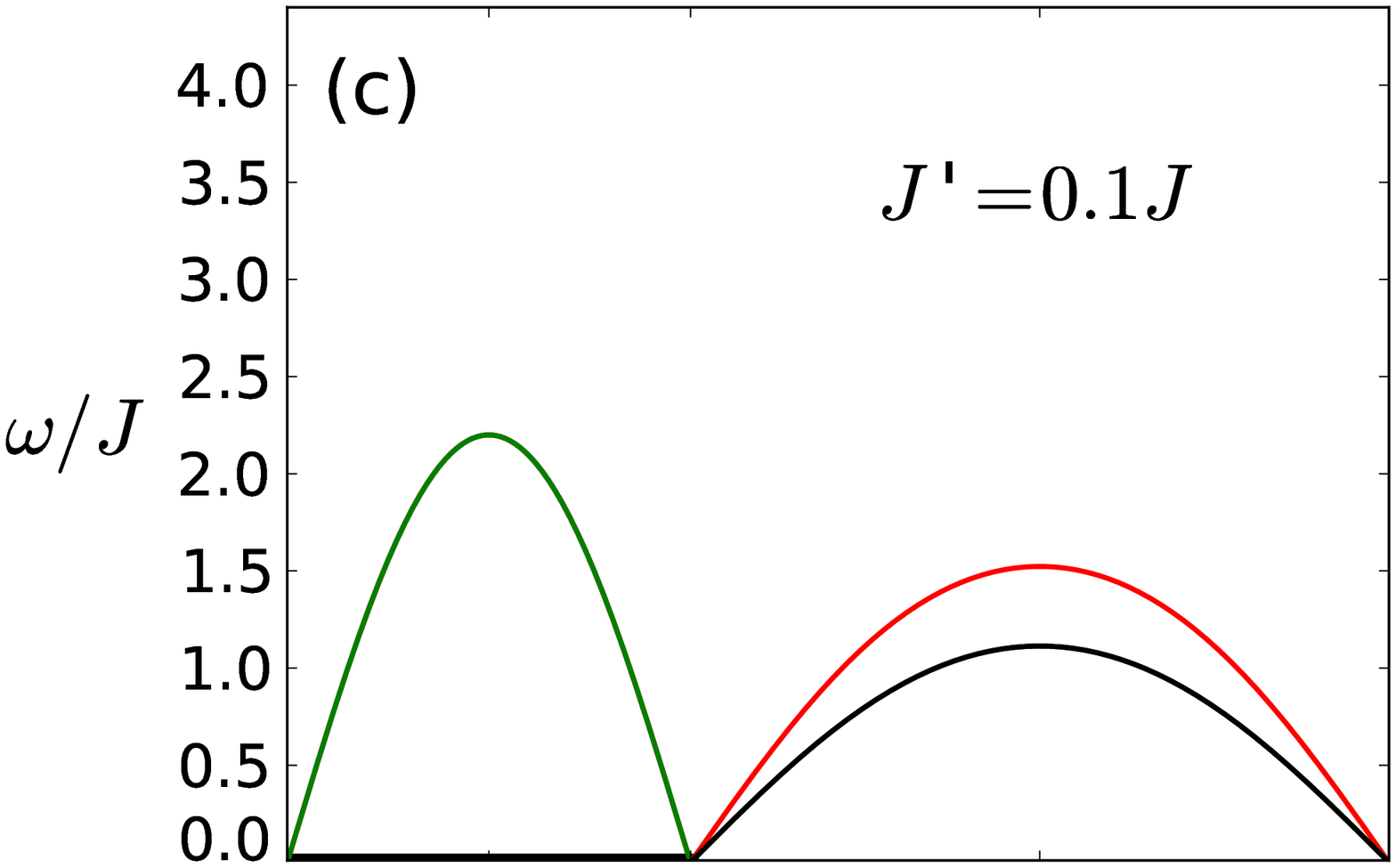}}
\subfloat{\label{subfig:J0.1K0.1}\includegraphics[width=50mm,height=45mm]{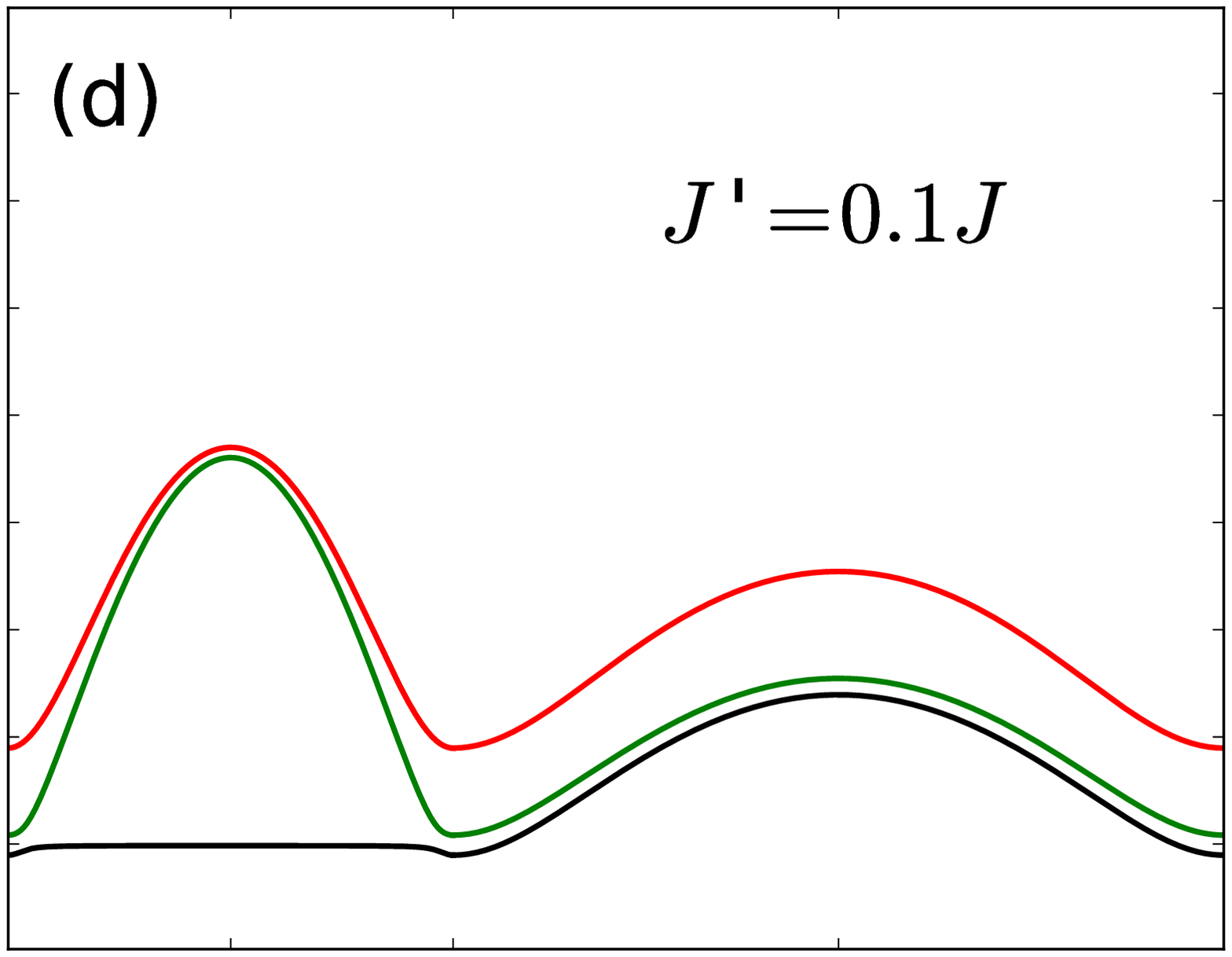}}
\\
\subfloat{\label{subfig:J1K0}\includegraphics[width=60mm,height=45mm]{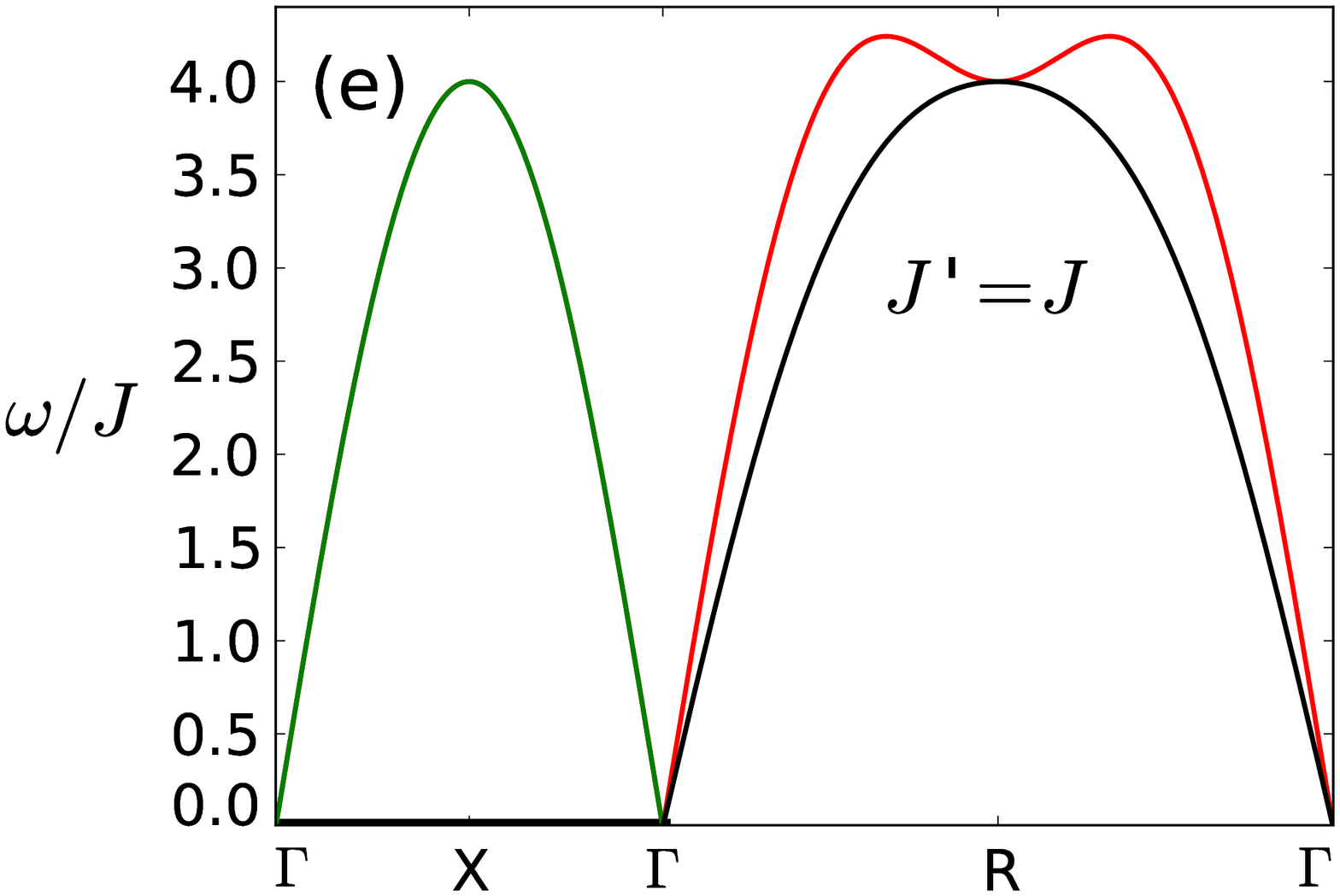}}
\subfloat{\label{subfig:J1K0.1}\includegraphics[width=50mm,height=45mm]{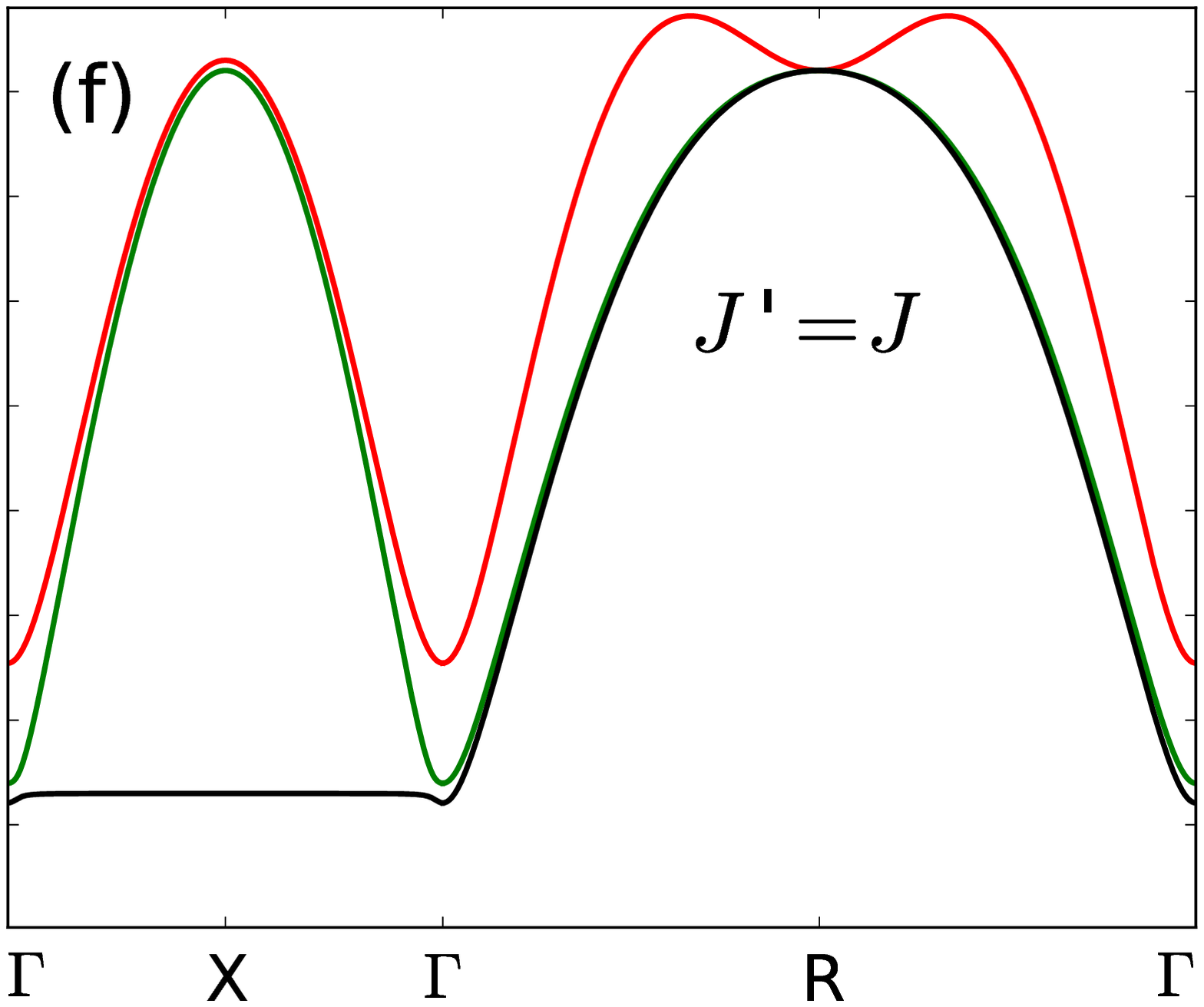}}
\caption{Spin wave modes along the $\Gamma$X and $\Gamma$R directions (a) $K=0, J'=0$, (b) $K=0.1J, J'=0$, (c) $K=0, J'=0.1J$, (d) $K=0.1J, J'=0.1J$, (e) $K=0, J'=J$, (f) $K=0.1J, J'=J$.}\label{Fig2}
\end{figure}

In the general case with $K>0$, the spin wave frequencies can only be obtained numerically. The ground state is no longer a planar configuration and the continuous degeneracies are removed and there are no zero modes. For values of $\mathbf{k}$ along the cubic axes, the lowest mode is almost dispersionless while the other two modes have strong dispersion. In all other wavevector directions, all three modes have strong dispersion.  Fig.~\ref{Fig2} shows the spin wave frequencies $\omega$ along the $\Gamma$X$(100)$ and $\Gamma$R$(111)$ directions for different values of the inter-plane coupling $J'$ and the cubic anisotropy $K$. For $K=0$, the effect of $J'$ is to stiffen the frequencies along $\Gamma$X and to remove the zero modes along $\Gamma$R. For $K>0$, the zero modes along both $\Gamma X$ and $\Gamma$R now have a substantial gap. Along the $\Gamma$X direction, there is a low frequency mode which is almost dispersionless (similar to the mode reported in Ref.~\cite{matan06}). Based on electronic structure calculations \cite{szunyogh2009}, the case $K=0.1J$ with $J' = J$ best represents IrMn$_3$. 

At the zone center $\mathbf{k}=\mathbf{0}$ we can obtain the leading behaviour of the three positive frequencies as a function of $K$
\begin{equation}
\begin{split}
\omega_1 \simeq \omega_2 \simeq&\ \sqrt{2(J+J')K} \\
\omega_3 \simeq&\ 2\sqrt{2(J+J')K} 
\end{split}
\end{equation}
Hence all modes have a gap for $K>0$ and although $\omega_1,\omega_2$ are degenerate to leading order in $K$, they become nondegenerate as $K$ increases. At the point R ($k_x=k_y=k_z=\pi/a $) and with $J'=J$, all three modes are degenerate for all $K$.

In our previous Monte Carlo simulations \cite{leblanc2013}, the sublattice magnetizations did not saturate at low $T$ and displayed evidence of degenerate spin configurations at $T=0$ for values of $K/J$ smaller than $\sim 0.06$. This behaviour is consistent with the presence of a small gap in the excitation spectrum. As $K$ increases, the gap increases
and the sublattices become fully saturated as $T=0$ is approached.

%\section{Neutron Scattering}
The effect of the cubic anisotropy on the spin excitation spectrum can be studied using inelastic magnetic neutron scattering. Here we consider the case of a single magnetic domain at zero temperature as above. Inelastic magnetic scattering is proportional to the dynamic structure factor 

\begin{equation}
S(\mathbf{q},\omega)= \sum_{m,n=x,y,z} S^{mn}(\mathbf{q},\omega)(\delta_{mn} - \hat{q}_m \hat{q}_n)
\end{equation}
where $S^{mn}(\mathbf{q},\omega)$ is the double Fourier transform of the correlation function $<S_i^m(0) S_j^n(t)>$ and can be calculated using the above results for the dispersion relations along with standard Green's functions techniques.\cite{MarshallandLovesey1971}
\begin{figure}[ht]
 \includegraphics[width=3in]{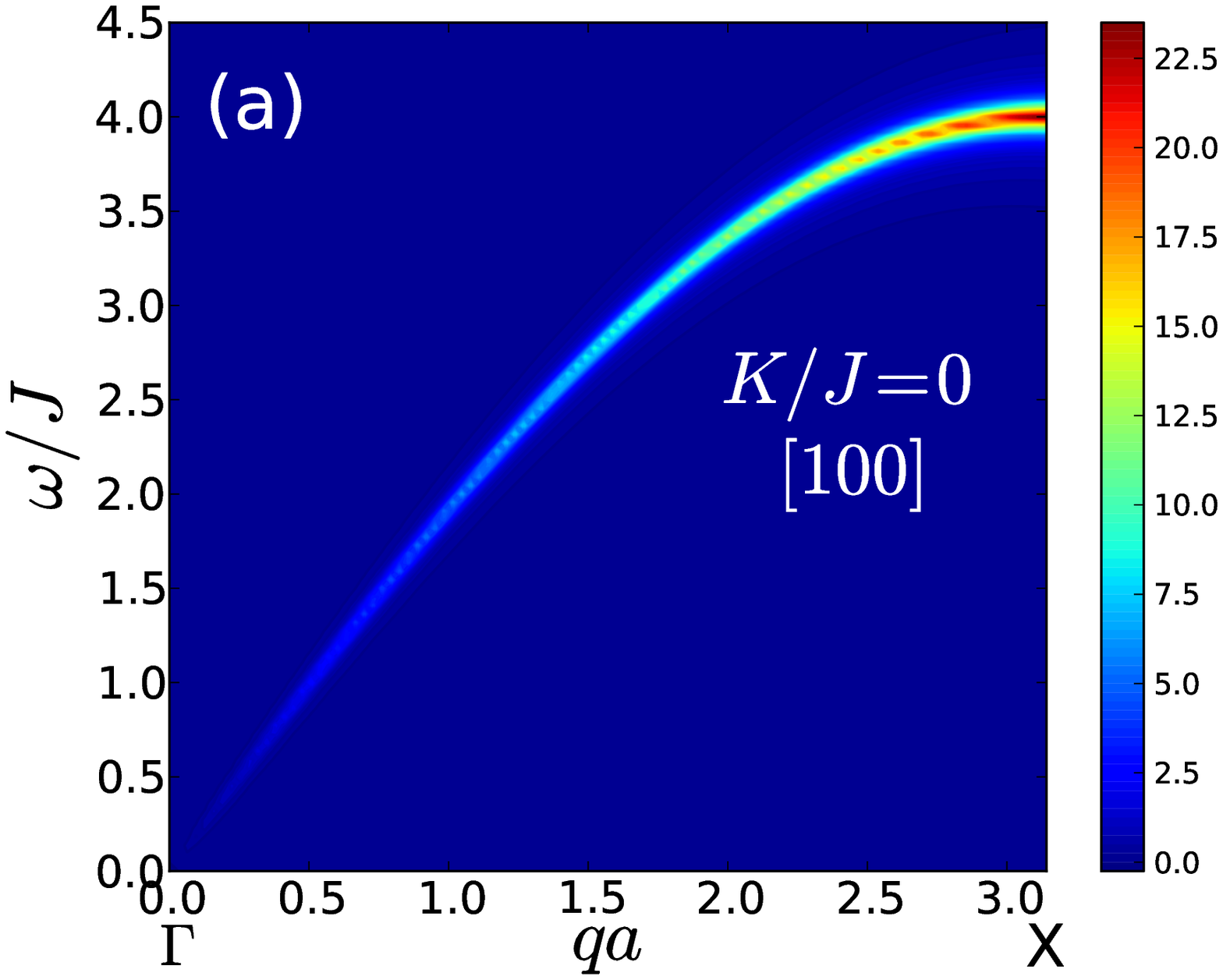}
 \includegraphics[width=3in]{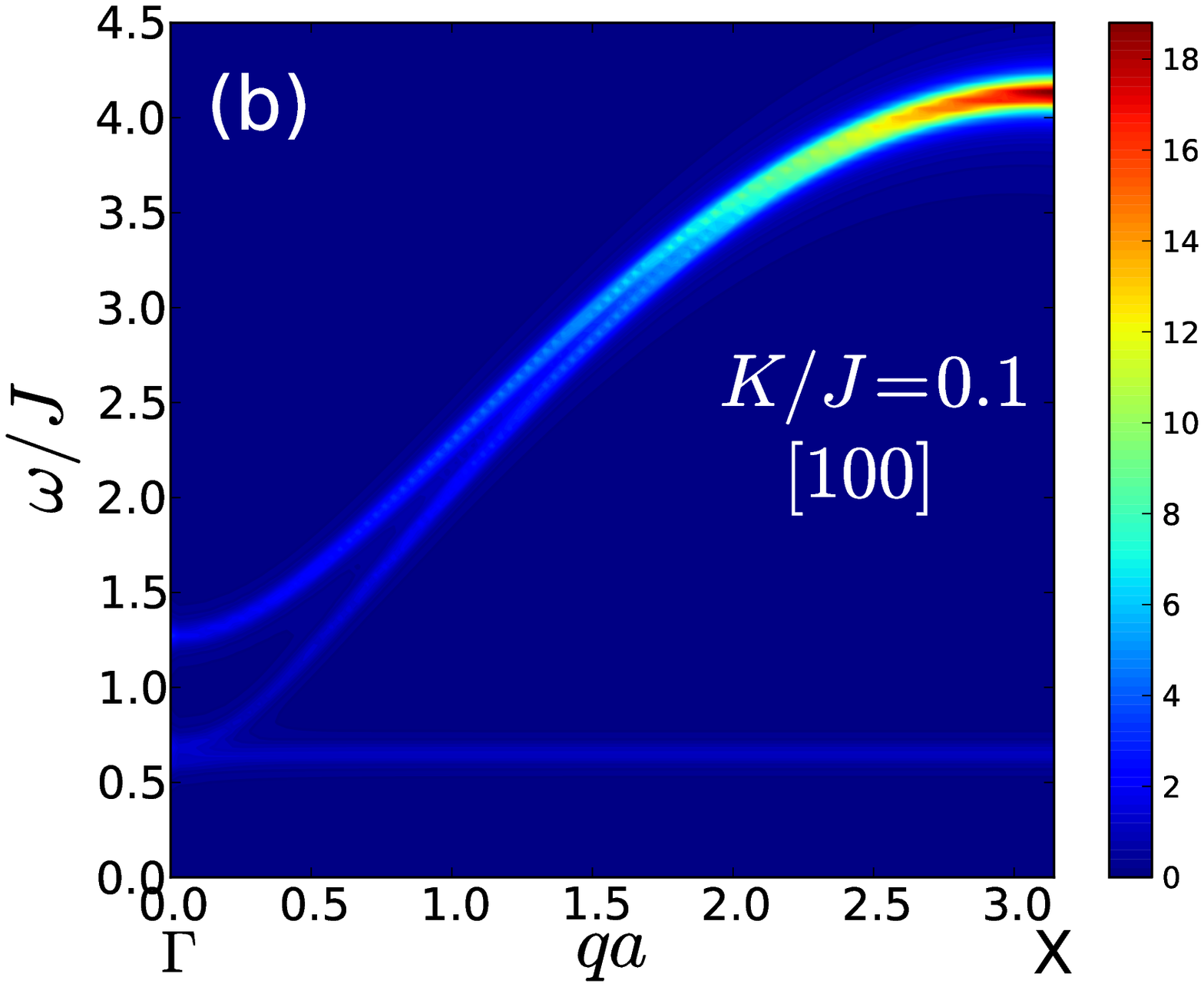}
 \includegraphics[width=3in]{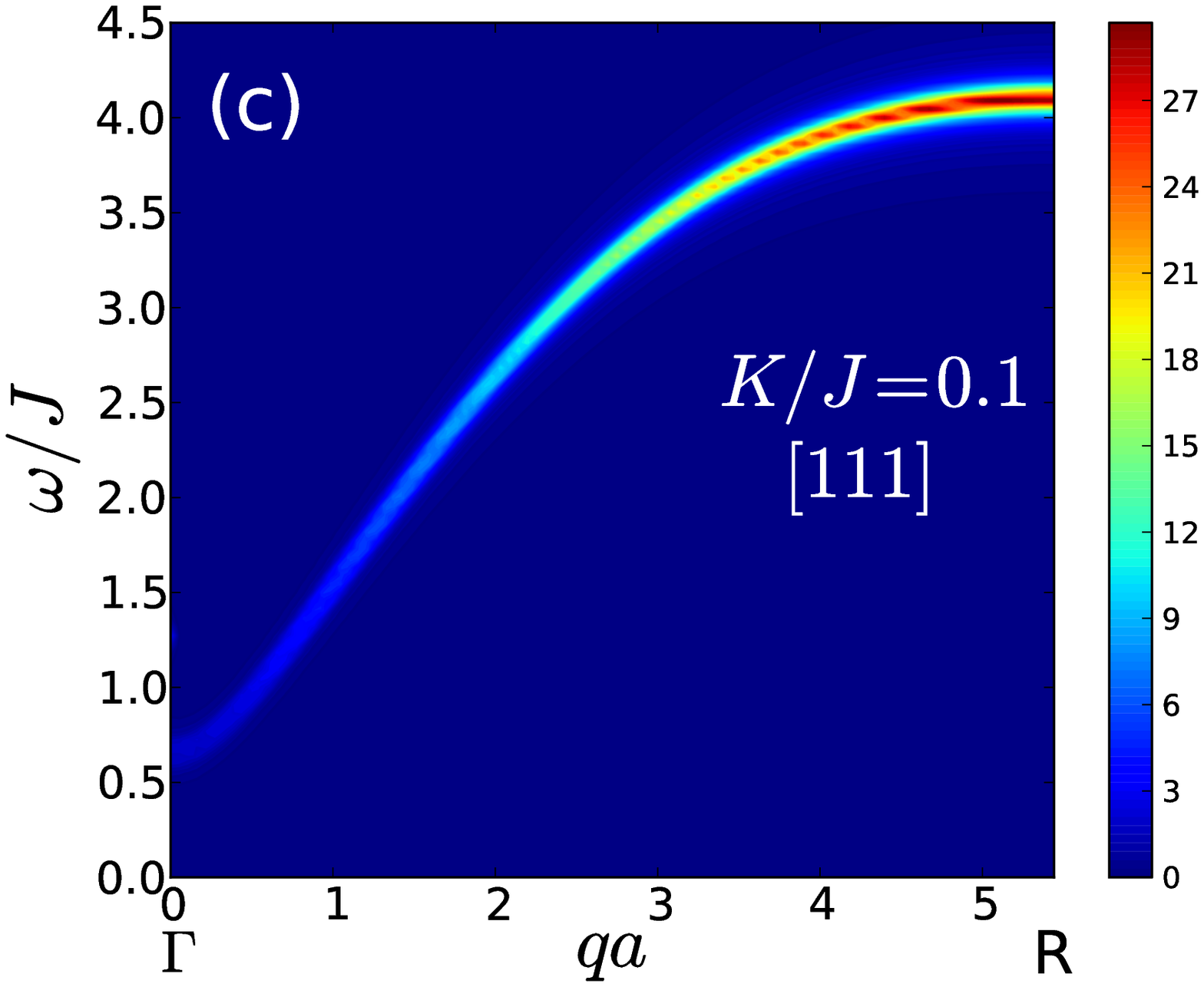}
 \includegraphics[width=3in]{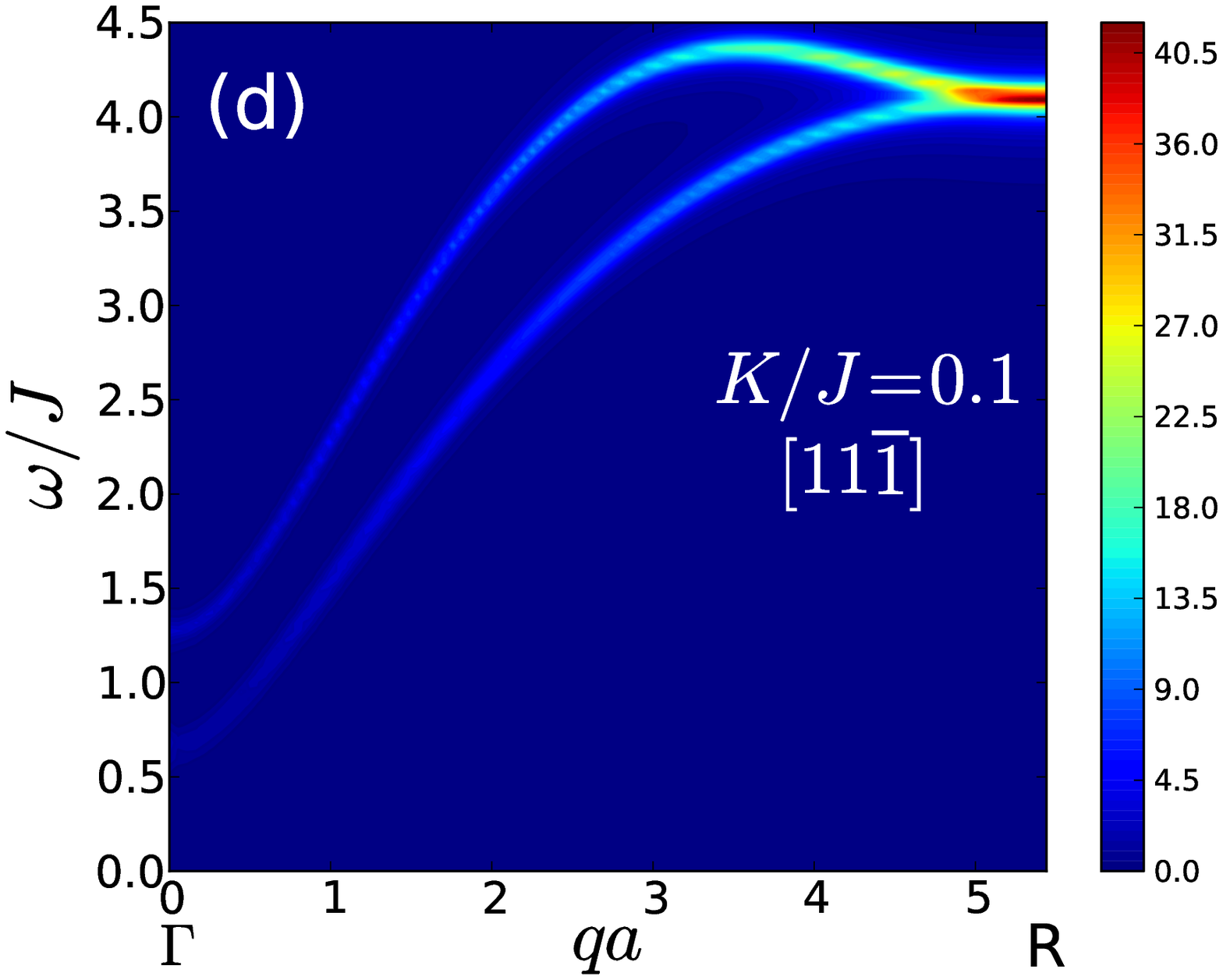}
 \caption{Relative magnitude of the inelastic scattering function $S(\mathbf{q},\omega)$ (side bar scale) assuming a single magnetic (111) domain with $J=J'=1$. (a) ${\bf q} \parallel [100]$ and $K$=0, (b)  ${\bf q} \parallel [100]$ and $K/J$=0.1, (c) ${\bf q} \parallel [111]$ and $K/J$=0.1, (d) ${\bf q} \parallel [11\overbar{1}]$ and $K/J$=0.1.}
\label{Fig3}
 \end{figure} 
 
Fig.~\ref{Fig3} shows $S(\mathbf{q},\omega)$ calculated with the assumption of a single magnetic (111) domain at fixed values of $\mathbf{q}$ along [100] and [111] directions with $J=J'=1$ for both $K=0$ and $K/J=0.1$. For ${\bf q} \parallel [100]$, Figs.~\ref{Fig3}(a) and (b) can be compared with Figs.~\ref{Fig2}(e) and (f), illustrating the appearance of the low frequency mode along $\Gamma$X and the splitting of the degeneracy of the higher frequency modes. However, along $[111]$ ($\Gamma$R), two modes are expected for $K=0$ and three modes for $K>0$, as in Figs.~\ref{Fig2}(e) and (f), but the highest frequency mode does not contribute to the scattering in either case as evident from Fig.~\ref{Fig3}(c). For the particular single domain considered here, this spin wave mode only has a dynamic magnetization parallel to the [111] axis and hence does not contribute to the scattering with the scattering vector $\mathbf{q}$ along $[111]$. However, this mode does contribute for $\mathbf{q}$ along the other equivalent [111] directions, as illustrated in Fig.~\ref{Fig3}(d).   Of particular note for all of the results shown in Fig.~\ref{Fig3} is that the intensity is expected to be relatively small near the zone center but is substantial larger at the zone boundary.  The form factor has not been included and would suppress the intensities near the zone boundary.
 
%\section{Summary and Conclusions}
The results of this work have demonstrated that the fcc kagome antiferromagnet is an example of the relatively rare phenomenon of macroscopic continuous degeneracy in 3D that gives rise to zero energy spin wave modes. Local cubic anisotropy is found to remove this degeneracy and introduce a gap in the spectrum. The lowest mode at small $K$ is almost dispersionless and has energy $\omega \sim 2\sqrt{JK}$ when $J'=J$, which is about $0.63 J$, assuming $K/J \approx 0.1$.  The electronic structure calculations on IrMn$_3$ provide the estimate $J \sim$ 40 meV \cite{szunyogh2009} giving $\omega \sim$ 25 meV. Anisotropy also induces a uniform magnetization in the [111] direction which could be utilized to stabilize a single-domain sample using field cooling techniques to better facilitate observation of these effects with inelastic neutron scattering experiments. These results support earlier Monte Carlo simulations which suggest that in the absence of anisotropy critical fluctuations drive the phase transition to be discontinuous but that it becomes continuous with the addition of anisotropy due to the removal of degeneracies. The model used in the present work can also serve as the foundation for further study of dynamic excitations associted with exchange bias phenomena in bilayer thin films that use IrMn$_3$.   

 %\section{Acknowledgments}
This work was supported by the Natural Sciences and Engineering Council (NSERC) of Canada, the Canada Foundation for Innovation (CFI), and Compute Canada.


\begin{thebibliography}{}

\bibitem{chalker1992} J. T. Chalker, P. C. W. Holdsworth, and E. F. Shender, Phys. Rev. Lett. \textbf{68}, 855 (1992).
\bibitem{harris1992} A. B. Harris, C. Kallin, and A. J. Berlinsky, Phys. Rev. B \textbf{45}, 2899 (1992).
\bibitem{schnabel2012} S. Schnabel and D. P. Landau, Phys. Rev. B \textbf{86}, 014413 (2012).
\bibitem{matan06} K. Matan, D. Grohol, D. G. Nocera, T. Yildirim, A. B. Harris, S. H. Lee, S. E. Nagler, and Y. S. Lee, Phys. Rev. Lett. \textbf{96}, 247201 (2006).
\bibitem{rastelli86} E. Rastelli and A. Tassi, J. Phys. C: Solid State Phys. \textbf{19}, L423 (1986); {\it ibid} \textbf{21}, 1003 (1988).
\bibitem{jansen86} A. P. J. Jansen, Phys. Rev. B \textbf{33}, 6352 (1986).
\bibitem{bramwell2001} S. T. Bramwell and M. J. P. Gingras, Science \textbf{294}, 1495 (2001).
\bibitem{zhitomirsky2012} M. E. Zhitomirsky, M. V. Gvozdikova, P. C. W. Holdsworth, and R. Moessner, Phys. Rev. Lett. \textbf{109}, 077204 (2012).
\bibitem{wong2013} A. W. C. Wong, Z. Hao, and M. J. P. Gingras, Phys. Rev. B \textbf{88}, 144402 (2013).
\bibitem{ross14} K. A. Ross, Y. Qiu, J. R. D. Copley, H. A. Dabkowska, and B. D. Gaulin, Phys. Rev. Lett. \textbf{112}, 057201 (2014).%arXiv:1401.1176v1(2014).
\bibitem{villain80} J. Villain, R. Bidaux, J. -P. Carton, and R. Conte, J. Phys. France \textbf{41}, 1263 (1980).
\bibitem{henley89} C. L. Henley, Phys. Rev. Lett. \textbf{62}, 2056 (1989).
\bibitem{shahbazi08} F. Shahbazi and S. Mortezapour, Phys. Rev. B \textbf{77}, 214420 (2008).
\bibitem{hemmati2012} V. Hemmati, M. L. Plumer, J. P. Whitehead, and B. W. Southern, Phys. Rev. B \textbf{86}, 104419 (2012).
\bibitem{leblanc2013} M. D. LeBlanc, M. L. Plumer, J. P. Whitehead, and B. W. Southern, Phys. Rev. B \textbf{88}, 094406 (2013). 
  
\bibitem{szunyogh2009} L. Szunyogh, B. Lazarovits, L. Udvardi, J. Jackson, and U. Nowak, Phys. Rev. B \textbf{79}, 020403(R) (2009).
\bibitem{berk1999} A. E. Berkowitz and K. Takano, J. Magn. Magn. Mater. \textbf{200}, 552 (1999); R. L. Stamps, J. Phys. D \textbf{33}, R247 (2000); M. Blamire and B. Hickey, Nat. Mater. \textbf{5}, 87 (2006).
\bibitem{ogrady2010} K. O'Grady, L. E. Fernandez-Outon, and G. Vallejo-Fernandez, J. Magn. Magn. Mater. \textbf{322}, 883 (2010).
\bibitem{tsunoda10} M. Tsunoda, H. Takahashi, T. Nakamura, C. Mitsumata, S. Isogami, and M. Takahashi, Appl. Phys. Lett. \textbf{97}, 072501 (2010).
\bibitem{kren66} Kr\'{e}n, G. K\'{a}d\'{a}r, L. P\'{a}l, J. S\'{o}lyom, and P. Szab\'{o}, Phys. Lett. \textbf{20}, 331 (1966);
E. Kr\'{e}n, G. K\'{a}d\'{a}r, L. P\'{a}l, J. S\'{o}lyom, P. Szab\'{o}, and T. Tarn\'{o}czi, Phys. Rev. \textbf{171}, 574 (1968);
A. Sakuma, R. Y. Umetsu, and K. Fukamichi, Phys. Rev. B \textbf{66}, 014432 (2002); T. Ikeda and Y. Tsunoda, J. Phys. Soc. Jpn. \textbf{72}, 2614 (2003).
\bibitem{tomeno1999} I. Tomeno, H. N. Fuke, H. Iwasaki, M. Sahashi, and Y. Tsunoda, J. Appl. Phys. \textbf{86}, 3853 (1999).

\bibitem{chen2014} H. Chen, Q. Niu, and A. H. MacDonald, Phys. Rev. Lett. \textbf{112}, 017205 (2014).
\bibitem{kohn2013} A. Kohn, A. Kovacs, R. Fan, G. J. McIntyre, R. C. C. Ward, and J. P. Goff, Sci. Rep. \textbf{3}, 2412 (2013).
\bibitem{yanes2013} R. Yanes, J. Jackson, L. Udvardi, L. Szunyogh, and U. Nowak, Phys. Rev. Lett. \textbf{111}, 217202 (2013).
\bibitem{reimers1993} Jan N. Reimers and A.J. Berlinsky, Phys. Rev. B \textbf(48), 9539 (1993).
\bibitem{torque} M. L. Plumer, J. Phys. C \textbf{17}, 4663 (1984); H. Tanaka, S. Teraoka, E. Kakehashi, K. Ito, and K. Nagata, J. Phys. Soc. Japan \textbf{57}, 3979 (1988).
\bibitem{morra1988} R. M. Morra, W. J. L. Buyers, R. L. Armstrong, and K. Hirakawa, Phys. Rev. B \textbf{38}, 543 (1988).


\bibitem{MarshallandLovesey1971} W. Marshall and S. W. Lovesey, {\it Theory of Thermal  Neutron Scattering} (Clarendon Press, Oxford, 1971).

\end{thebibliography}
\end{document}